\documentclass[aps,twocolumn,showpacs,showkeys,floatfix]{revtex4}

\usepackage{graphicx}
\usepackage{dcolumn}
\usepackage{bm}

\begin{document}


\title{X-ray diffraction from shock-loaded polycrystals}

\date{February 19, 2006; revised July 4, 2007}


\author{Damian C. Swift}
\email{dswift@lanl.gov}
\affiliation{%
   P-24 Plasma Physics, Los Alamos National Laboratory,
   MS~E526, Los Alamos, New Mexico 87545, USA
}

\begin{abstract}
X-ray diffraction was demonstrated from shock-compressed polycrystalline metal 
on nanosecond time scales.
Laser ablation was used to induce shock waves in polycrystalline foils of Be, 
25 to 125\,$\mu$m thick.
A second laser pulse was used to generate a plasma x-ray source by irradiation 
of a Ti foil.
The x-ray source was collimated to produce a beam of controllable diameter, and 
the beam was directed at the Be sample.
X-rays were diffracted from the sample, and detected using films and x-ray 
streak cameras.
The diffraction angle was observed to change with shock pressure.
The diffraction angles were consistent with the uniaxial (elastic)
and isotropic (plastic) compressions expected for the loading conditions
used.
Polycrystalline diffraction will be used to measure the response of 
the crystal lattice to high shock pressures and through phase changes.
\end{abstract}


\maketitle

\section{Introduction}
Shock wave experiments and static presses such as diamond anvil cells are the 
principal techniques for investigating the
behavior of materials at high pressures \cite{Emerets96}.
Shock wave experiments are particular important when investigating, 
and developing models
of, the response of materials to dynamic loading such as the shocks themselves.
One of the key measurements in static presses is diffraction, which provides 
direct observation of the compression and strain of the crystal lattice, 
and can be used to detect and identify phase transitions.
These measurements are important comparisons with theoretical predictions such 
as electronic structure calculations.

Diffraction in static presses is routinely performed from samples which are 
single crystals or polycrystalline ensembles \cite{Emerets96}.
In shock wave experiments, the usual experimental measurements include the 
speed of the shock wave and the material speed behind the shock.
Such observations can be used to infer the mechanical equation of state and 
some aspects of plastic flow.
Diffraction data from shocked samples are potentially extremely valuable as 
direct measurements of the elastic compression of the crystal lattice, the 
onset of plastic flow, phase transitions, and the density of crystal defects 
such as dislocations.
The width of the diffraction lines may allow the temperature of the shocked 
material to be deduced, which is a long-standing problem in shock physics.

X-ray diffraction has been performed in shock wave experiments, using single 
crystal samples loaded by projectile impact \cite{Johnson70,Rigg01} and by 
laser ablation \cite{Loveridge01}.
The length of time for which the high pressure state persists is short, 
typically for microseconds in projectile impact experiments and nanoseconds or 
less for ablative loading.
It is difficult to generate x-rays with sufficient
power to acquire a diffraction pattern on these time scales, and to 
synchronize the x-rays with the shock.
X-ray diffraction patterns have been obtained by integrating the signal as 
samples are shocked repetitively \cite{RosePetruck99,Kishimura02}, but these 
experiments were performed using single crystals loaded by ablation from 
sub-picosecond laser pulses.
It is not practical to perform large number of repeat measurements 
with current projectile guns and high energy lasers, and it is in any case 
highly desirable to collect the diffraction pattern during a single, 
well-characterized shock event.

\begin{figure}
\begin{center}
\includegraphics[scale=0.6]{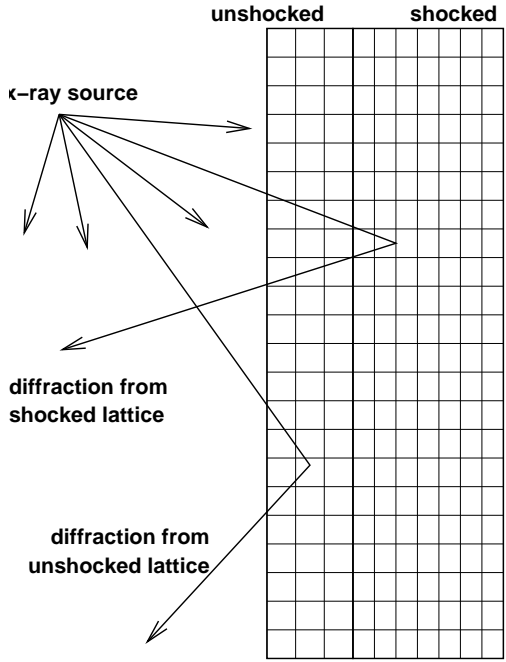}
\end{center}
\caption{Schematic of transient x-ray diffraction from a single crystal sample,
   illustrating diffraction from different points at different compressions.}
\label{fig:xtaltxd}
\end{figure}

Diffraction from shocked polycrystalline material has not been reported 
previously, but is desirable in several respects.
Using a single crystal, the point at which diffraction occurs moves across
the sample as the lattice spacing changes.
For high compressions, it can be difficult to ensure that both the unshocked
and shocked lines can be accommodated over the finite width of the shocked
region (Fig.~\ref{fig:xtaltxd}).
In contrast, polycrystalline diffraction is best arranged from a single
point (or small area) of the sample, which does not move with compression.
Another compelling reason is that, even if a sample is initially a single 
crystal, if it undergoes a phase transition then the daughter phase may be 
polycrystalline.
For many materials, polycrystalline samples are far easier to obtain than are 
single crystals, and it is desirable to study the properties of materials in 
the form in which they are used in engineering applications.
It is also interesting to study the effect of grain size and sample texture on 
properties such as the dynamics of phase changes.

\begin{figure}
\begin{center}
\includegraphics[scale=0.18]{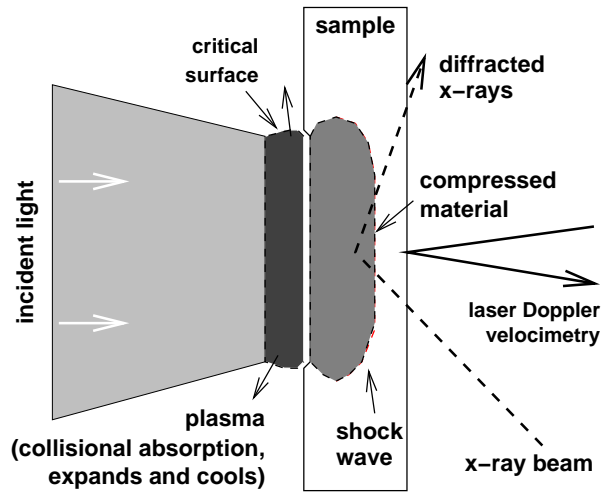}
\end{center}
\caption{Schematic of shock generation by laser ablation.}
\label{fig:lasershock}
\end{figure}

We have performed x-ray diffraction experiments on shock-loaded single crystals 
of a variety
of materials, notably Si \cite{Loveridge01,Swift_sieos_01} and Be 
\cite{Swift_betxd_sccm01}.  
Be-based alloys are of technological interest as the deuterium-tritium
fuel capsule for inertial confinement fusion (ICF) 
\cite{Wilson98,Bradley99,Lindl98}.
Our experiments on Be investigate the anisotropy of plastic
response under dynamic loading, and phase transitions including melting.
As well as their technological significance, time-resolved x-ray diffraction 
data from these experiments
are valuable in understanding plasticity in hexagonal metals and the dynamics
of melting.
Shock loading was induced by laser ablation (Fig.~\ref{fig:lasershock}): 
the loading history is often less
simple and less well-characterized than is the case for projectile impact
experiments, but laser-induced shocks are far easier to synchronize with
laser-powered x-ray sources.  They are also more appropriate for the length and 
time
scales of interest for ICF: $\sim$100\,$\mu$m and 1\,ns.
We have developed an accurate predictive capability to
relate the pressure history of the shock to the irradiance history of the
laser pulse \cite{Swift_elements_04,Swift_alloys_04}.  Radiation hydrodynamics
simulations can be used to understand the loading history, and to adjust the
temporal structure of the pulse to induce a constant drive pressure and hence
a better-characterized shock.

\section{Laser ablation experiments}
These experiments were performed at the TRIDENT laser facility at Los Alamos 
National Laboratory.
TRIDENT is based around
a Nd:glass laser with a fundamental wavelength of 1054\,nm.
To improve the coupling to matter and reduce the risk of damage to the laser 
from backscatter of the fundamental, for these experiments the beams were 
frequency-doubled to 527\,nm.
Each sample was loaded by irradiation using the one beam of the laser;
pressure was generated by ablation of material from the irradiated surface
(Fig.~\ref{fig:lasershock}).

Because of the experimental configuration used (specifically, 
the design of the nose of the x-ray streak cameras), it was difficult to
acquire velocimetry and x-ray diffraction data at the same time.
Two types of experiment were performed: a series with no diffraction 
measurements in which 
the velocity history of the opposite surface (across the thin direction)
was measured using line-imaging Doppler velocimetry for a wide range of 
irradiance histories \cite{Swift_bevisar_01,Swift_besim_04}, and a series of 
diffraction measurements in which the velocity history was measured on only 
some shots.
The irradiance history was measured in all cases, so the loading history could 
be calculated using radiation hydrodynamics and continuum mechanics simulations.

\begin{figure}
\begin{center}
\includegraphics[scale=0.9]{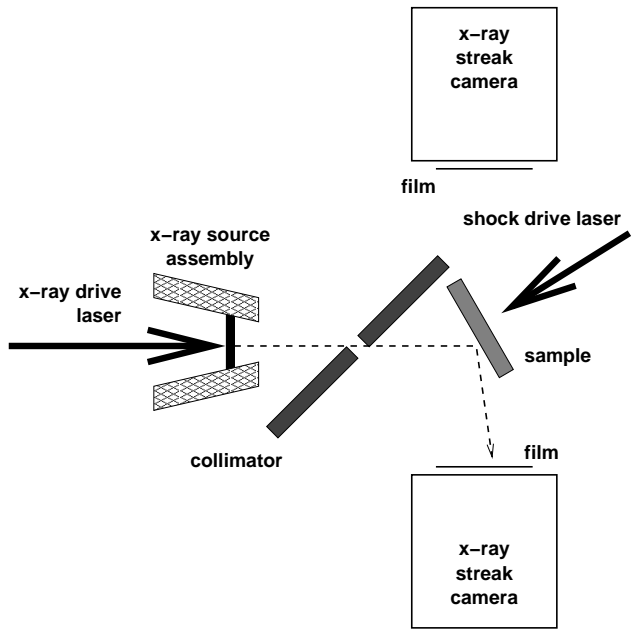}
\end{center}
\caption{Schematic of experimental layout (not to scale).
   The slit in each x-ray streak camera was in the plane
   of the diagram.}
\label{fig:exptlayout}
\end{figure}

X-rays were generated by irradiating a thin metal foil with a second
laser pulse, tightly-focused, to produce a hot plasma.
The resulting x-rays were collimated to illuminate the center of the sample
with a narrow beam.
In our previous experiments with single crystals 
\cite{Loveridge01,Swift_sieos_01,Swift_betxd_sccm01},
diffraction produced cones of x-rays with a virtual apex at the x-ray source,
reflected by the diffracting plane in the sample.
For a collimated x-ray beam and a polycrystalline sample,
diffraction produces a series of cones whose axis is the incident x-ray
beam and apex is the diffracting point on the sample \cite{McKie74}.
Films and x-ray streak cameras were positioned to record photons diffracted 
around 90$^\circ$ from the incident beam, where powder diffraction from Be 
would produce a signal.
The slit of the stream camera was parallel to the x-ray beam,
so a given change in powder diffraction angle would produce the greatest
change in line position.
For polycrystal diffraction from thin samples, 
the angle of the sample with respect to the x-rays is not important.
In these experiments, the sample was mounted normal to the drive beam,
which was at 136$^\circ$ from the x-ray beam.
(Fig.~\ref{fig:exptlayout}.)

The laser beam used to induce the shock driving beam and the laser beam used
to generate x-rays were both driven from the same master oscillator,
so they shared the same duration and un-amplified pulse shape,
though the energies were controlled independently.
The pulse shape was chosen to optimize the shock drive (keeping the pressure
as constant as possible in time);
the energy of the x-ray generating pulse was chosen to optimize 
the spectrum and intensity of x-rays.
The timing of the x-rays was defined with respect to the shock drive,
by changing the path length of the x-ray driving beam.
Typical delays were $\sim$2\,ns.

\subsection{Sample material}
\begin{figure}
\begin{center}
\includegraphics[scale=0.55]{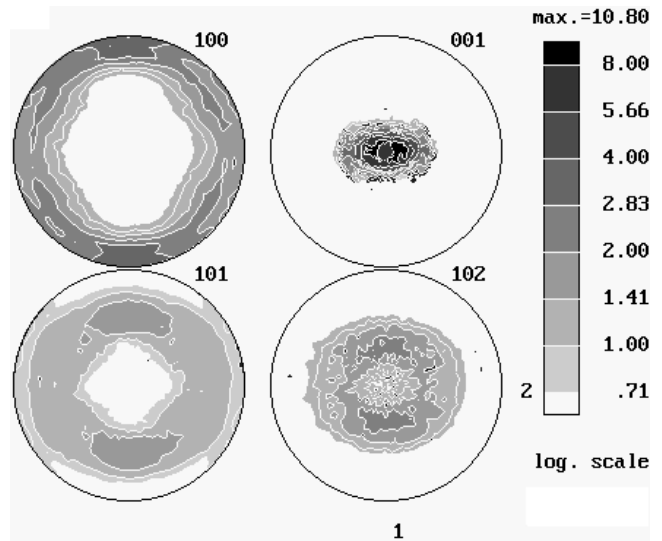}
\end{center}
\caption{Orientation maps of crystal planes in foils 125\,$\mu$m thick.
   Each circular figure is a stereographic projection, showing a greyscale map
   of the distribution of normals to one plane.
   The center of the circle is the normal to the sample; the circumference
   is the set of directions lying in the plane of the sample.}
\label{fig:xtalorient}
\end{figure}

Be foils were obtained from Goodfellow Inc, of 99.8\%\ purity
(the maximum supplied).
These were rolled foils, with significant texture.
The texture was investigated by obtaining x-ray orientation maps of a couple
of sample foils.
The scattering efficiency of Be is low, so the thicker foils
(125\,$\mu$m) were used in these measurements.
The foils were found to be oriented preferentially so that normals to
the $(0001)$ planes were distributed around the normal to the foil,
equivalent to a rocking curve a couple of tens of degrees wide.
The $(10\bar 10)$ planes were distributed with normals in the plane of
the sample, though with significant variation in this azimuthal distribution,
reflecting the directions in which the foil was rolled.
(Fig.~\ref{fig:xtalorient}.)

\subsection{Target assembly}
\begin{figure}
\begin{center}
\includegraphics[scale=0.75]{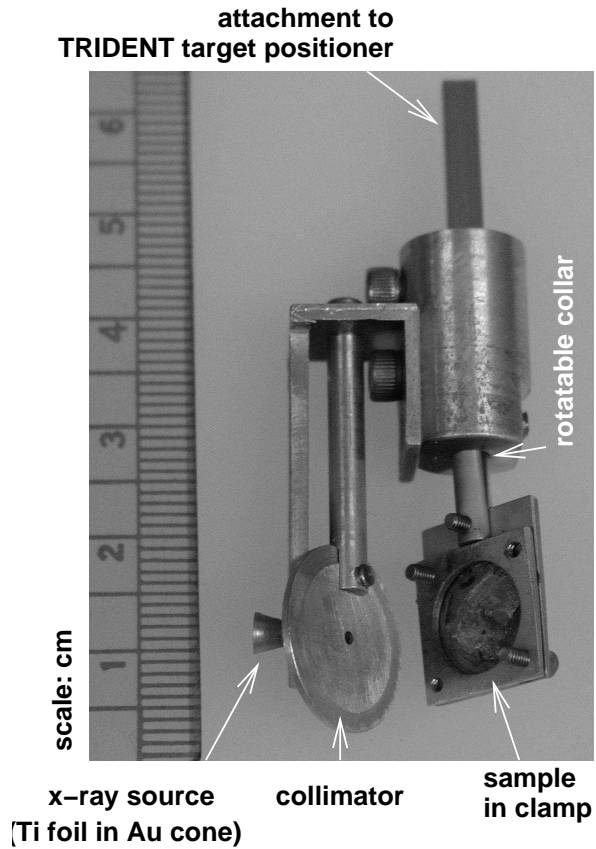}
\end{center}
\caption{Re-usable clamp-type target holder (`Ortiz holder').}
\label{fig:holder}
\end{figure}

A re-usable holder was used to locate the sample and x-ray source reproducibly 
for each shot (Fig.~\ref{fig:holder}).
The sample assembly was clamped by the edges, the holder having
apertures for the drive and velocimetry beams.
The holder was designed for use with a wide sample assembly, i.e.
a large aperture for the sample, so that the clamp-type arrangement
used to hold the sample would be well out of the region illuminated by
the x-rays and thus unlikely to increase the x-ray background by fluorescence.
Thus either a wide sample was used($\sim$10\,mm across), or the sample
was attached to a retaining mount (e.g. a washer) of a material which
would not cause problems with x-ray fluorescence.
For the Be experiments, pieces were cut which were large enough to fill 
the full aperture in the holder.

\subsection{Shock-driving beam}
TRIDENT was operated in nanosecond mode, in which each pulse comprised
up to 13 consecutive elements whose intensity was controlled separately.
Using all elements, the drive pulse would be 2.5\,ns long.
For the time-resolved x-ray diffraction experiments reported here, the pulse 
length was shorter
(fewer elements used) to optimize the efficiency of energy conversion to x-rays.
The pulses generated were approximately square, with a slight ramp
to induce a more constant pressure.
The shock-driving beam was
focused through a Fresnel zone plate to produce a uniform irradiance
over a spot 5\,mm in diameter \cite{Swift_lice_05}.
The pulse energy and intensity history were recorded on each shot.

The power history delivered by each laser beam was measured in each experiment 
by directing the light reflected from uncoated glass plates in the beam path 
onto a fast photodiode and a calorimeter.
The spatial distribution of laser in the target plane had been measured 
separately \cite{Swift_lice_05}, and was used to convert the power history to 
the irradiance history.
The calorimeter reading had been calibrated previously against a calorimeter
placed at the center of the target chamber.
The uncertainty in energy was of the order of 1\,J.

\subsection{Velocity history}
For the supporting experiments without diffraction, 
the velocity history of the free surface of the sample was measured 
by time-resolved
laser Doppler velocimetry, using a system of the 'velocity interferometry for 
surfaces of any reflectivity' (VISAR) type \cite{Barker72}.
Line VISAR illumination was provided by a pulsed Nd:YAG laser, operating
at 1319\,nm wavelength with output frequency-doubled to 660\,nm.
The laser pulse was around 50\,ns long.
Line VISAR fringes were recorded using an optical streak camera.
The fringe constant of the VISAR was computed 
from the thickness of the delay element (3" of SF6 glass).
The dispersion of the glass was determined at the wavelength of the
probe laser by fitting a straight line to points sampling the variation
of refractive index with wavelength.
The fringe constant deduced was 806\,m/s at 660\,nm wavelength.

\subsection{X-ray diffraction}
\begin{figure}
\begin{center}
\includegraphics[scale=0.7]{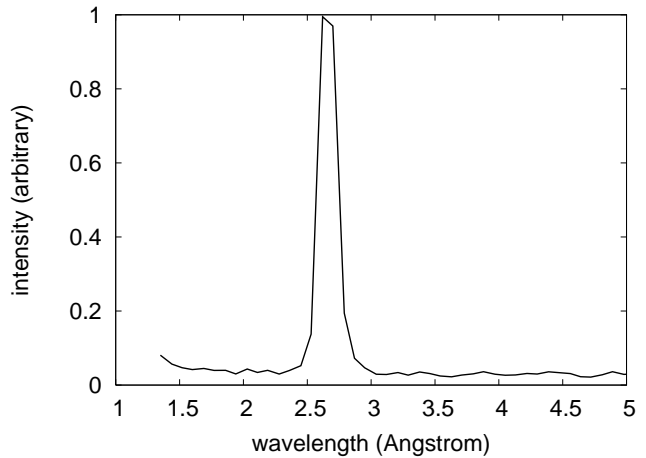}
\end{center}
\caption{X-ray spectrum from laser-heated Ti foil.}
\label{fig:spectrum}
\end{figure}

The plasma x-ray source used one beam from TRIDENT, focused tightly
($\sim$150\,$\mu$m diameter spot) on a metal foil $\sim$10\,$\mu$m thick
to generate a hot plasma.
The plasma emitted He-$\alpha$ radiation from recombination of the ionized
atoms.
The x-ray spectrum was measured with a grating spectrometer, recording on
Kodak Corp.
DEF film and was thus a time-average over the duration of x-ray emission.
The spectrum was
essentially monochromatic for the range and precision of diffraction angles
in these experiments (Fig.~\ref{fig:spectrum}).
The source material was chosen according to the anticipated x-ray diffraction
angles: Ti (He-$\alpha$ at 2.61\,\AA\ or 4.75\,keV)
for use with Be at compressions of up to a few tens of percent.

The plasma source emitted x-rays roughly isotropically into $4\pi$ steradians.
The collimator was an Al disk, 1\,mm thick and 15\,mm in diameter.
collimator.
A hole 1\,mm in diameter was drilled through the center to define the beam size.
This rather large size was chosen so that a signal would be more likely to be
observed in early trials.  The collimator was twisted with respect to
the x-ray axis in subsequent experiments to present a narrower (non-circular) 
aperture.

The orientation of the films and cameras was chosen so that diffraction from
Be $\{0002\}$ planes would be within the field of view.
The energy in the laser beam heating the x-ray source was 
typically 150 to 200\,J over 1.0\,ns for a Ti source.
The x-ray source foil was mounted inside a truncated gold cone, which prevented 
direct irradiation of the x-ray detectors by the source.
The emission spectrum was recorded using a transmission grating and DEF film.

\begin{figure}
\begin{center}
\includegraphics[scale=0.7]{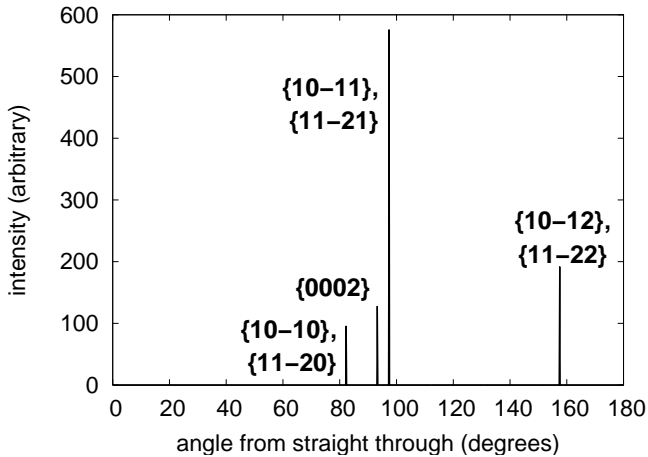}
\end{center}
\caption{Simulated powder pattern for polycrystalline Be.}
\label{fig:bepowder}
\end{figure}

\begin{figure}
\begin{center}
\includegraphics[scale=0.7]{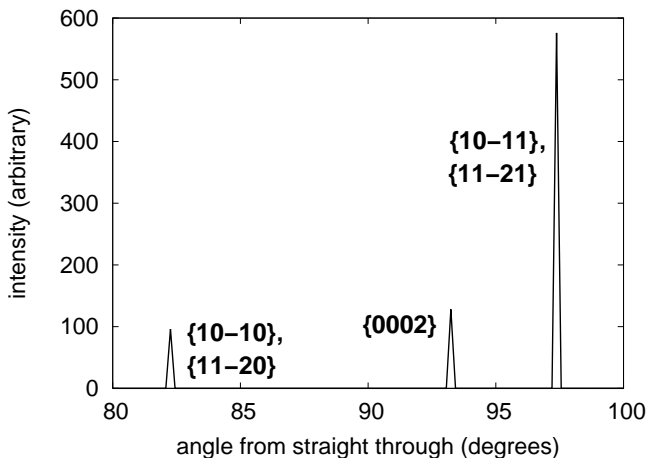}
\end{center}
\caption{Simulated powder pattern for polycrystalline Be
   (range of angles captured by detectors).}
\label{fig:bepowder1}
\end{figure}

X-rays from the plasma were incident on the sample material,
and diffracted from individual grains according to the Bragg condition.
The attenuation length of $\sim$5\,keV x-rays in Be is several
hundred microns, compared with sample thicknesses of a few tens of microns.
Thus the diffraction signals were averages through the entire thickness of
each sample, in general comprising material in different states of compression.
If compressed material was present -- e.g. behind the shock wave -- 
during the backlighter pulse, the x-rays were diffracted at a different angle.
If the grains in the sample were oriented isotropically, the 
powder diffraction pattern obtained with x-rays of the wavelength used in
these experiments would comprise a cluster of three lines within $10^\circ$
of $90^\circ$ from the undiffracted beam
(Figs~\ref{fig:bepowder} and \ref{fig:bepowder1}).

Diffracted x-rays were recorded on DEF films (time-integrating)
and Kentech x-ray streak cameras.
The light signal from the streak tube was amplified with an image intensifier
before recording on TMAX optical film.
In some cases, signals were observed on the x-ray streak cameras
but not on the time-integrating films placed in front of the cameras,
because the cameras were much more sensitive than the film.
CsI photocathodes were used in the x-ray streak cameras.
CsI is relatively sensitive but has a short life in experiments of this
type: in some experiments, the CsI was damaged and produced spurious signals.
The streak camera slits each subtended an angle of $12^\circ$ at the target.
The time-integrating films subtended an angle of $20^\circ$, constrained by
the inside diameter of the cylindrical camera snout.

Filter foils were added to control the signal level, 
in front of the film pack, and between the film pack
and the photocathode of the x-ray streak camera.
125\,$\mu$m of Be was used in all cases, with around
25\,$\mu$m of Ti, the exact thickness adjusted for best signal.

\section{Loading histories}
The loading histories were calculated using a method described previously
\cite{Swift_elements_04,Swift_alloys_04,Swift_lice_05}.
Radiation hydrodynamics simulations were used to predict the loading history
applied to the sample, with the sample material treated as a fluid.
The predicted loading history was then used as a time-dependent
pressure boundary condition for continuum mechanics simulations
in which elastic, plastic, and damage models were included for the sample
material.

The experiments with surface velocimetry were used as the primary diagnostic
to validate the simulations. 
These wave profile measurements yielded the timing and amplitude
for any elastic precursor wave and the plastic shock,
and the detailed pressure history.

\subsection{Radiation hydrodynamics}
As shown previously \cite{Swift_elements_04},
ablative laser loading on nanosecond time scales with irradiances
$\sim$0.1 to 10\,PW/m$^2$ can be simulated
accurately by treating the ablation region 
($\sim$1\,$\mu$m wide of the solid sample and the resulting plasma cloud)
with three temperature hydrodynamics (ions, electrons,
and radiation), thermal conduction and radiation diffusion,
and laser absorption through the electrical conductivity.
The remainder of the condensed phase was treated using non-radiative
continuum mechanics, i.e. with an equation of state (EOS) and
a constitutive model.
Some specific processes excluded in this plasma model, i.e. negligible
in the regime considered, include resonant
absorption, transport of the resulting hot electrons, and generation,
transport, and deposition of Bremsstrahlung x-rays.

Simulations were performed using the
HYADES radiation hydrocode, version 01.05.11 \cite{HYADES}.
This hydrocode used a 1D Lagrangian discretization of the material,
and leapfrog time integration,
and did not include material strength in the solid sample.
Shock waves were stabilized using artificial viscosity.
The EOS and opacity were represented using tabular models
from the SESAME database \cite{SESAME}.
Conductivities for laser deposition and heat conduction were
calculated using the Thomas-Fermi ionization model
\cite{HYADES,Zeldovich66};
this was found previously to be reasonably accurate for direct drive
shock simulations on samples of a wide range of atomic numbers
\cite{Swift_elements_04}.
The flux limiter was set to 0.03 of the free stream value -
a common choice for simulations of this type \cite{Dendy93}.

The initial spatial mesh was set up to be expanding,
to allow adequate resolution of the material to be ablated.
Moving away from the sample surface, adjacent cells were expanded by 5\%.
The cell closest to the surface was 5\,nm wide.
Previous sensitivity studies had demonstrated that this resolution was adequate
for direct drive simulations in this regime of irradiance and time scale
\cite{Swift_elements_04}.
Where possible, simulations were performed using the irradiance history measured
on each shot to infer the precise loading history applied to
each sample.
The exceptions were experiments in which the photodiode measuring drive
history was saturated; in these cases, the shape was assumed to be the
same as for shots fired shortly before or after, and the amplitude was
scaled to give the correct total energy.
The raw photodiode record was converted to irradiance by scaling so that
the integral under the curve matched the measured pulse energy,
and dividing by the area of the focal spot.
The raw record contained high frequency fluctuations, giving negative
apparent values of the irradiance when the signal was low.
For the simulations, the inferred irradiance history was reduced to
a piecewise linear variation reproducing the principal features to
a few percent in instantaneous irradiance.

\subsection{Continuum mechanics}
Continuum mechanics simulations were used to predict the propagation and
evolution of the loading history through the thickness of the sample,
taking account of the constitutive properties (elasticity and strength)
of the material.
Simulations were performed using the
LAGC1D hydrocode, version 5.2 \cite{LAGC1D}.
This hydrocode used a 1D Lagrangian discretization of the material,
and predictor-corrector time integration.
Shock waves were stabilized using artificial viscosity.

\section{Results}
These experiments were performed at the TRIDENT laser facility 
during July-August 2000, July-August 2001, and August 2002 
(`Flying Pig 2' shot series).
For the pulses nominally 1.0 and 1.8\,ns long, the laser pulse comprised 
6 and 9 elements respectively. 
No-shock test shots were performed to check for the signal level
in static diffraction with the nanosecond laser-heated x-ray source,
and to optimize camera gain and filtering.
Dynamic shots were performed with shock pressures between 10 and 50\,GPa.
Some shots were performed with similar shock conditions to check
reproducibility.
On a few shots, the irradiance history of the shock-driving beam exhibited
significant, undesired variations.
On the remainder, the irradiance was close to the desired history, and
the shock-driving pressure was constant to within around 10\%.

\subsection{Static trials}
\begin{figure}
\begin{center}
\includegraphics[scale=0.6]{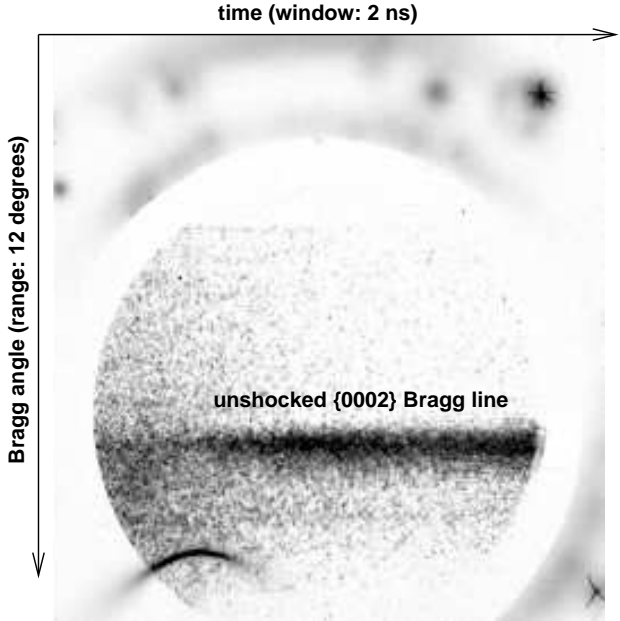}
\end{center}
\caption{Example x-ray streak camera record from a static (unshocked) trial:
   TRIDENT shot 13664.
   The spots and star-shaped markings at the outside were caused by arcing 
   from the pulsed high voltage in the streak tube.
   The circular image intensifier is clearly evident.}
\label{fig:staticbraggstreak}
\end{figure}

\begin{figure}
\begin{center}
\includegraphics[scale=0.45]{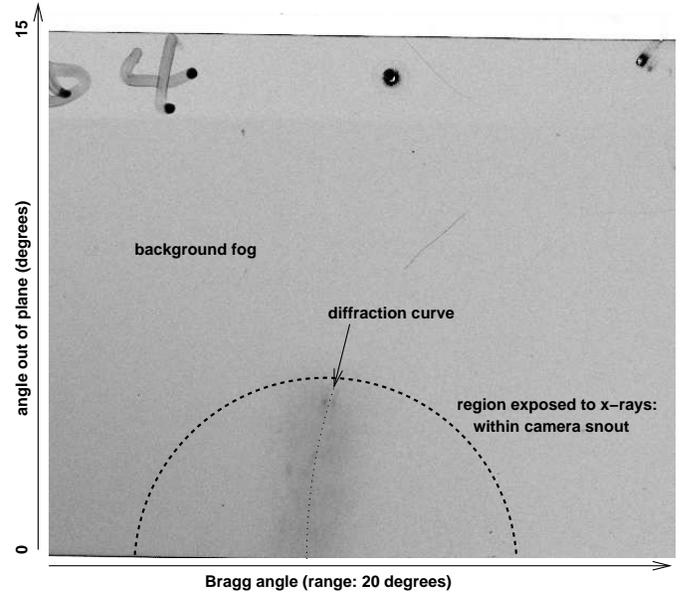}
\end{center}
\caption{Example x-ray film record from a static (unshocked) trial:
   TRIDENT shot 15004.}
\label{fig:staticbraggfilm}
\end{figure}

With no shock drive, a clear diffraction signal was observed from 
the $\{0002\}$ planes, as a line on the time-resolved streak camera
record (Fig.~\ref{fig:staticbraggstreak}, 
and as an arc on the time-integrated film (Fig.~\ref{fig:staticbraggfilm}).
The arc was mottled, presumably corresponding to diffraction from 
specific favorably-oriented grains in the sample.
The arc was the visible part of the conic section between the 
cone of diffracted x-rays and the film plane; the rest of the curve
was obscured by the streak camera snout, which was in front of the
cassette for the time-integrating film.
Apart from the desired diffraction line, there was no significant
x-ray exposure above the background fog level of the film,
indicating that scatter or fluorescence from Al and steel components
in the target holder was not a problem.
The breadth of the arc, and the width of the time-resolved line,
varied with the width of the beam passed by the collimator.
The incident beam was adjusted so that the doublet in the x-ray source
spectrum was visible in the x-ray streak record.

The target chamber geometry constrained the location of the `Laue' streak camera
and the static film mount attached to its snout so that unshocked lines were 
barely visible at the extreme ends of the static field of view.

\subsection{Shock trials}
\begin{figure}
\begin{center}
\includegraphics[scale=0.45]{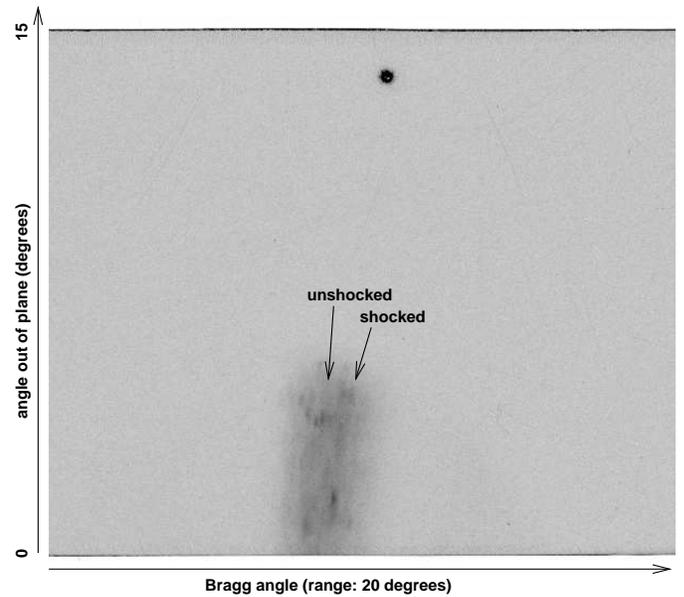}
\end{center}
\caption{Example x-ray film record from a shocked sample:
   TRIDENT shot 15002.}
\label{fig:braggfilm}
\end{figure}

\begin{figure}
\begin{center}
\includegraphics[scale=0.55]{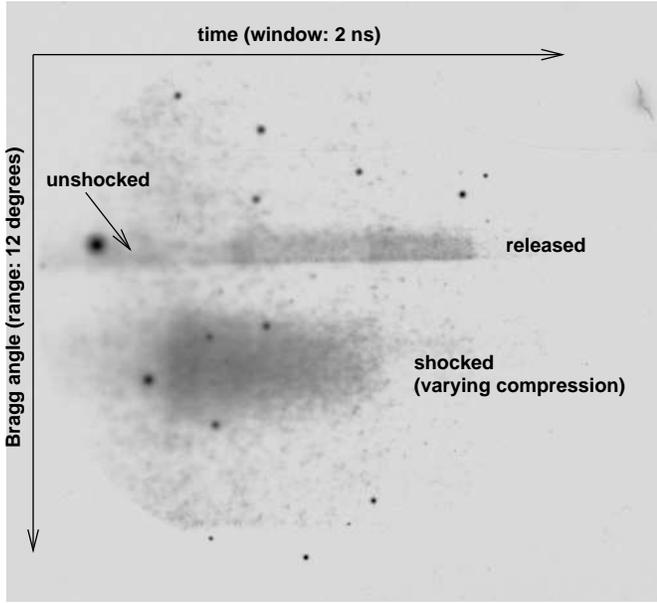}
\end{center}
\caption{Example x-ray streak camera record from a shocked sample:
   TRIDENT shot 15002.}
\label{fig:braggstreak}
\end{figure}

\begin{figure}
\begin{center}
\includegraphics[scale=0.7]{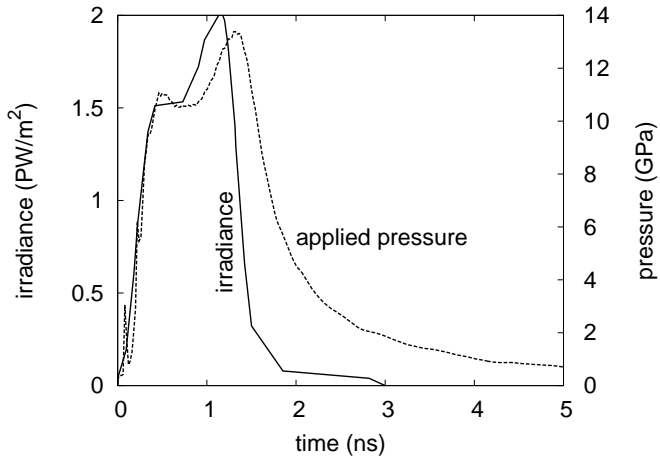}
\end{center}
\caption{Irradiance and drive pressure history from example dynamic loading experiment
   (TRIDENT shot 15002).}
\label{fig:drivehist}
\end{figure}

With a shock drive, one or more displaced arcs appeared on the 
time-integrated film
(Fig.~\ref{fig:braggfilm}),
and one or more additional lines appeared on the streak camera record
(Fig.~\ref{fig:braggstreak}).
Clearest results were obtained for shock pressures around 10-20\,GPa
(Fig.~\ref{fig:drivehist}).
At higher pressures, the diffraction lines passed out of the field of view
of the detectors.
In general, the x-ray streak record showed a sharp unshocked line,
growing weaker as the shock compressed more material;
diffraction from shocked material, less sharp than the
unshocked line presumably because of the slightly varying shock pressure;
and a sharp line from released material with a lattice parameter
slightly greater than that of the unshocked material.
The displacement of the $\{0002\}$ Bragg reflection
was consistent between the time-resolved and time-integrated measurements,
but could be measured more accurately from the time-resolved record.

\begin{figure}
\begin{center}
\includegraphics[scale=0.7]{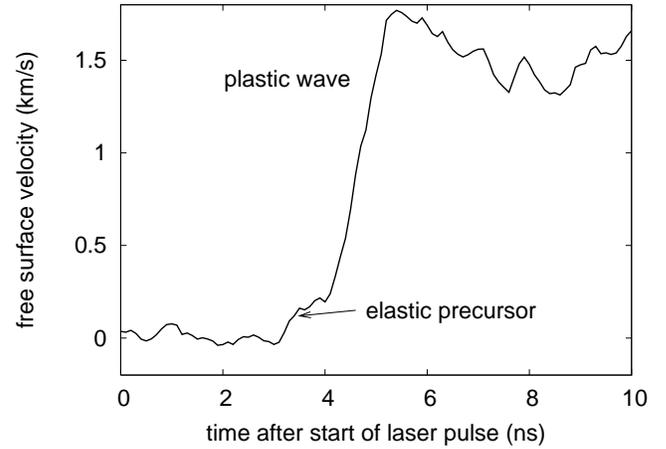}
\end{center}
\caption{Example free surface velocity history from a shocked sample:
   TRIDENT shot 12182.}
\label{fig:linevis}
\end{figure}

One complication in interpreting the shocked signal was that material
compressed uniaxially in the elastic precursor wave had a similar lattice
parameter to material compressed more isotropically in the plastic wave.
The amplitude of the elastic wave was observed clearly in the line VISAR
measurements of free surface velocity history (Fig.~\ref{fig:linevis}).

\section{Interpretation of diffraction lines}
\begin{figure}
\begin{center}
\includegraphics[scale=0.7]{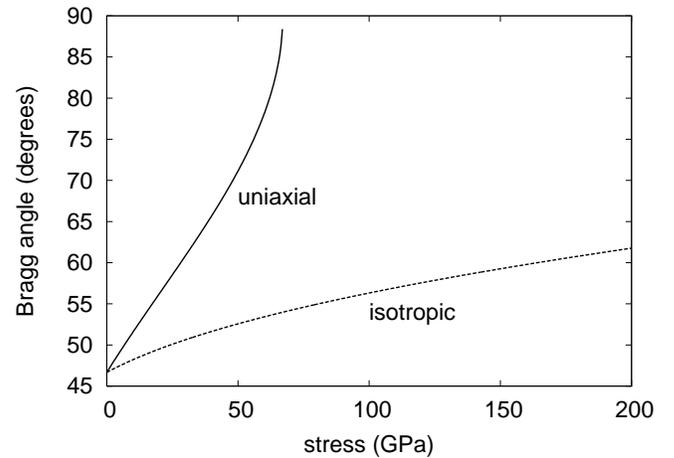}
\end{center}
\caption{Dependence of $(0002)$ Bragg angle in Be on shock pressure,
   for limiting assumptions about the deformation symmetry of the lattice.}
\label{fig:hugtheta}
\end{figure}

The interpretation of a single diffraction line in the absence of other
data depends on the assumed orientation of the plane with respect to the
shock, and on the assumed symmetry of deformation of the lattice.
The extreme cases of lattice deformation are pure uniaxial or pure isotropic,
and this assumption leads to significant differences in the compression
or stress inferred from a given change in diffraction angle 
(Fig.~\ref{fig:hugtheta}).

Pure uniaxial lattice deformation occurs in elastic waves only;
pure isotropic deformation occurs when the flow stress is negligible.
When the drive pressure is not high enough to overdrive the elastic
precursor, the diffraction signal would in principle consist of
uniaxial compression to the elastic strain corresponding to the flow stress,
and partially-isotropic compression to the highest pressure.
Plastic flow acts to return the elastic strain to the flow stress,
so the high pressure state generally includes a uniaxial component of this
magnitude rather than pure isotropic compression.
(The isotropic stress state is generally not an isotropic strain state for
a single crystal, as the elastic constants generally vary with orientation.)
When the drive pressure is high enough to overdrive the elastic wave,
the high-pressure state is of course all that would be observed.
If the elastic wave reaches an impedance mismatch at the opposite surface
of the sample (e.g. and usually a free surface), it will reverberate between
the surface and the plastic shock, creating regions of intermediate strain.

When interpreting time-resolved and time-integrated records,
the detailed signal depends on
the timing of the x-ray pulse with respect to the propagation of the
different waves and the range of penetration of the x-rays in the sample.
For Be, x-rays of the energy used here can penetrate hundreds of
microns, so the diffraction signal was integrated over all positions
through the thickness of the sample.

The time-resolved Bragg records typically showed a strong doublet near the
center, again from $(0002)$ planes.
In experiments with shock loading, the time-resolved Bragg records also showed 
a broad signal corresponding to lattice compression.
The degree of compression can be converted to a uniaxial stress or
an isotropic pressure by assuming a degree of uniaxiality.
Assuming uniaxial strain around the $[0001]$ direction, the range of
angles in for example shot 15002
implied a flow stress in the range 3 to 8\,GPa, which is consistent
with values deduced from velocimetry.
Assuming isotropic compression, the range of angles implied a shock pressure
in the range 7 to 20\,GPa, bracketing the expected shock pressure but with
a much larger range than expected for any individual experiment.
Unfortunately, the Bragg angles expected for uniaxial and isotropic compression
in this range were fairly similar: the observed record is certainly a
superposition of both types of deformation.
This ambiguity should be absent in the higher-compression experiments
needed to investigate shock-induced melting, as the Bragg angles would be
significantly different and the elastic wave may be overdriven in many cases;
a wider field of view would be needed to detect Bragg angles over the
necessary range of compressions.
The ambiguity would also be less in experiments on softer materials.
In shot 15012, the shocked signal moved with time, possibly indicating
lattice relaxation i.e. the transition from uniaxial to isotropic compression.
The drive pressure was predicted to be relatively constant (compared with
the change in lattice parameter); it seems possible that the signal
has captured the decay of a very strong elastic precursor ($\sim10$\,GPa) 
at early times.

The time-integrated Laue records typically showed two faint signals close to
the edges of the field of view, i.e. around $15^\circ$ apart.
The position and spacing are consistent with the $\{10\bar 10\}$ and
$\{11\bar 20\}$ lines at one edge and the 
$\{10\bar 11\}$ and $\{11\bar 21\}$ lines at the other.
Presumably the $\{0002\}$ lines did not appear -- and the lines above did
not appear on the Bragg record -- because of the texture of the rolled foils.
The position and low signal levels made it difficult to extract any useful
distribution of angles.

\section{Discussion}
Experiments were performed in which Be foils were loaded
by laser ablation, and the distortion of the grains was monitored
with {\it in situ} x-ray diffraction.
X-ray powder patterns were clearly visible on the time-integrated (film) record
positioned to capture reflections from $\{0002\}$ planes,
and there was evidence of lines at the edges of the opposite time-integrating
record, consistent with diffraction from 
$\{10\bar 10\}$, $\{11\bar 20\}$, $\{10\bar 11\}$, and $\{11\bar 21\}$ planes.
In some experiments, the x-ray streak cameras also recorded diffraction lines,
but some apparent lines may have been caused by camera faults.
The records clearly included signals from shocked and unshocked Be.
The deviations in Bragg angle were consistent with strains anticipated from
the ablative loading applied, and the range of flow stress deduced from
surface velocimetry.
In these initial experiments, the accuracy and separation of the changing 
diffraction lines was not sufficient to be used in isolation -- velocimetry
is essential, 
and a wider angular coverage for x-ray detection is highly desirable.
The measurements did unambiguously demonstrate diffraction from
shocked, polycrystalline material.

\section*{Acknowledgments}
We would like to thank the TRIDENT staff, particularly Randy Johnson, 
Tom Hurry, Tom Ortiz, Fred Archuleta, Nathan Okamoto, and Ray Gonzales,
for their hard work on the experiments.
Bernie Carpenter (P-24) and Kathy Gallegos (Bechtel) scanned the films.
Jon Larsen (Cascade Applied Sciences, Inc.) gave advice on the
use of the radiation hydrocode HYADES.

Funding was provided by Allan Hauer and Steve Batha of
the LANL Program Office for the
National Nuclear Security Administration's Campaign 10,
Inertial Confinement Fusion.
The work was performed under the auspices of
the U.S. Department of Energy under contract W-7405-ENG-36.

\end{document}